\documentclass[a4paper,11pt]{article}
\usepackage{pos}

\title{Machine learning approaches for parameter reweighting in MC samples of top quark production in CMS}

\author*[1]{Valentina Guglielmi}
\note{For the CMS Collaboration.}

\affiliation{Deutsches Elektronen-Synchrotron (DESY),\\
  Notkestrasse 85, Hamburg, Germany}

\emailAdd{valentina.guglielmi@desy.de}
\newcommand{\rb}{\ensuremath{r_\text{b}}}
\newcommand{\xb}{\ensuremath{x_\text{b}}}
\newcommand{\ptB}{\ensuremath{p_\text{T}^\text{B}}}
\newcommand{\pB}{\ensuremath{p_\text{B}}}

\newcommand{\ttbar}{\textrm{\ensuremath{\mathrm{t\bar{t}}}}}
\newcommand{\mt}{\ensuremath{m_\text{t}}}
\newcommand{\mW}{\ensuremath{m_\text{W}}}
\newcommand{\pt}{\ensuremath{p_\text{{T}}}}
\newcommand{\hdamp}{\ensuremath{h_\text{{damp}}}}

\abstract{In particle physics, Monte Carlo (MC) event generators are needed to compare theory to the measured data. Many MC samples have to be generated to account for theoretical systematic uncertainties, at a significant computational cost. Therefore, the MC statistic becomes a limiting factor for most measurements and the significant computational cost of these programs a bottleneck in most physics analyses. In this contribution, the Deep neural network using Classification for Tuning and Reweighting (DCTR) approach is evaluated for the reweighting of two systematic uncertainties in MC simulations of top quark pair production within the CMS experiment. DCTR is a method, based on a Deep Neural Network (DNN) technique, to reweight simulations to different model parameters by using the full kinematic information in the event. This methodology avoids the need for simulating the detector response multiple times by incorporating the relevant variations in a single sample.}

\FullConference{%
  European Physical Society, EPS2023\\
  20-25 August, 2023\\
  Hamburg, Germany
}

\begin{document}
\maketitle

\section{Introduction}
\label{sec:Intro}
In particle physics, Monte Carlo (MC) event generators are essential for comparing theoretical predictions with experimental data. Generating numerous MC samples is necessary to account for theoretical uncertainties, which comes at a significant computational cost. As a result, the limitations of MC statistics hinder many measurements, and the high computational expense poses a bottleneck for physics analyses. For instance, in a recent study on top quark-antiquark pair production, the main source of uncertainty in the mass of the top quark came from the MC statistics of the samples used for systematic estimation \cite{paper}. To address this issue, reweighting the MC samples can be a solution. This approach involves generating only a sample with nominal values, and then variations are obtained by reweighting this nominal sample. By doing so, the need for simulating the detector response multiple times is eliminated, reducing the MC statistics and computational cost.
In contrast to traditional reweighting, which compares distributions in specific bins at a truth level, Machine Learning (ML) reweighting offers a more flexible approach. Standard reweighting is limited by the choice of binning and the number of input dimensions. ML reweighting, using a neural network, does not have such limitations. It can utilize all event information as input, improving reweighting precision. Additionally, it allows for the simultaneous reweighting of multiple MC parameters, considering their correlations. This project introduces Deep Neural Network (DNN) using Classification for Tuning and Reweighting (DCTR) \cite{DCTR} and tests its performance in two scenarios \cite{DPS}: discrete reweighting of the $\hdamp$ variation in parton-level Powheg HVQ events and continuous reweighting of a b-fragmentation function parameter in particle-level Pythia8 events.

\begin{figure}[htbp]
\centering
\includegraphics[width=0.9\textwidth]{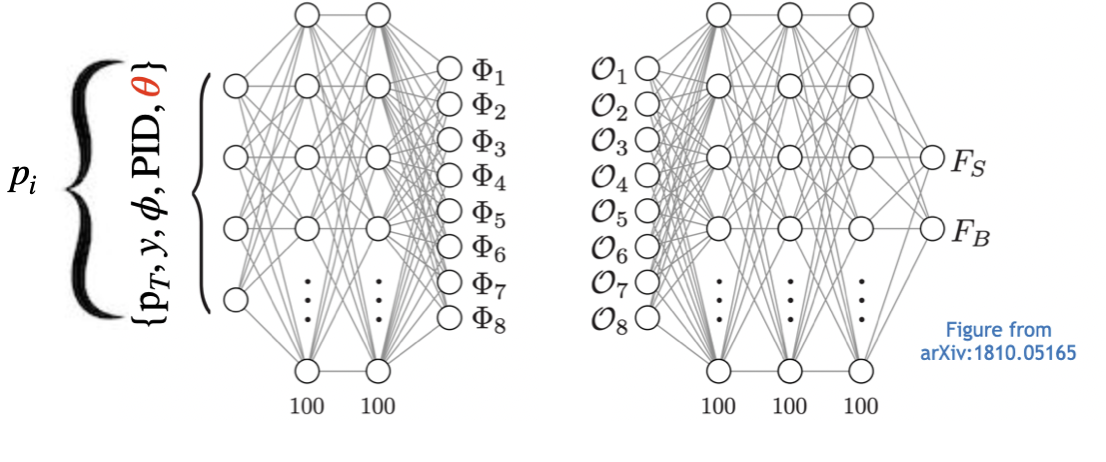}
\caption{The PFN architecture used in the DCTR method is depicted. 
It parametrizes the per-particle mapping $\Phi$ (on the left) and the function F (on the right), shown for the case of a latent space of dimension $l = 8$. The latent observable is $\ensuremath{O_\mathrm{a}} = \sum_{i}\ensuremath{\Phi_\mathrm{n}}(\pt, \ensuremath{y_\mathrm{i}}, \ensuremath{\phi_\mathrm{i}}, \ensuremath{m_\mathrm{i}}, PID\ensuremath{_\mathrm{i}}$)~\cite{set}.}
\label{fig:NN2}
\end{figure}

\section{$h_{damp}$ dicrete reweighting}
\label{sec:Hdamp}
\begin{figure}[htbp]
\centering
\includegraphics[width=0.45\textwidth]{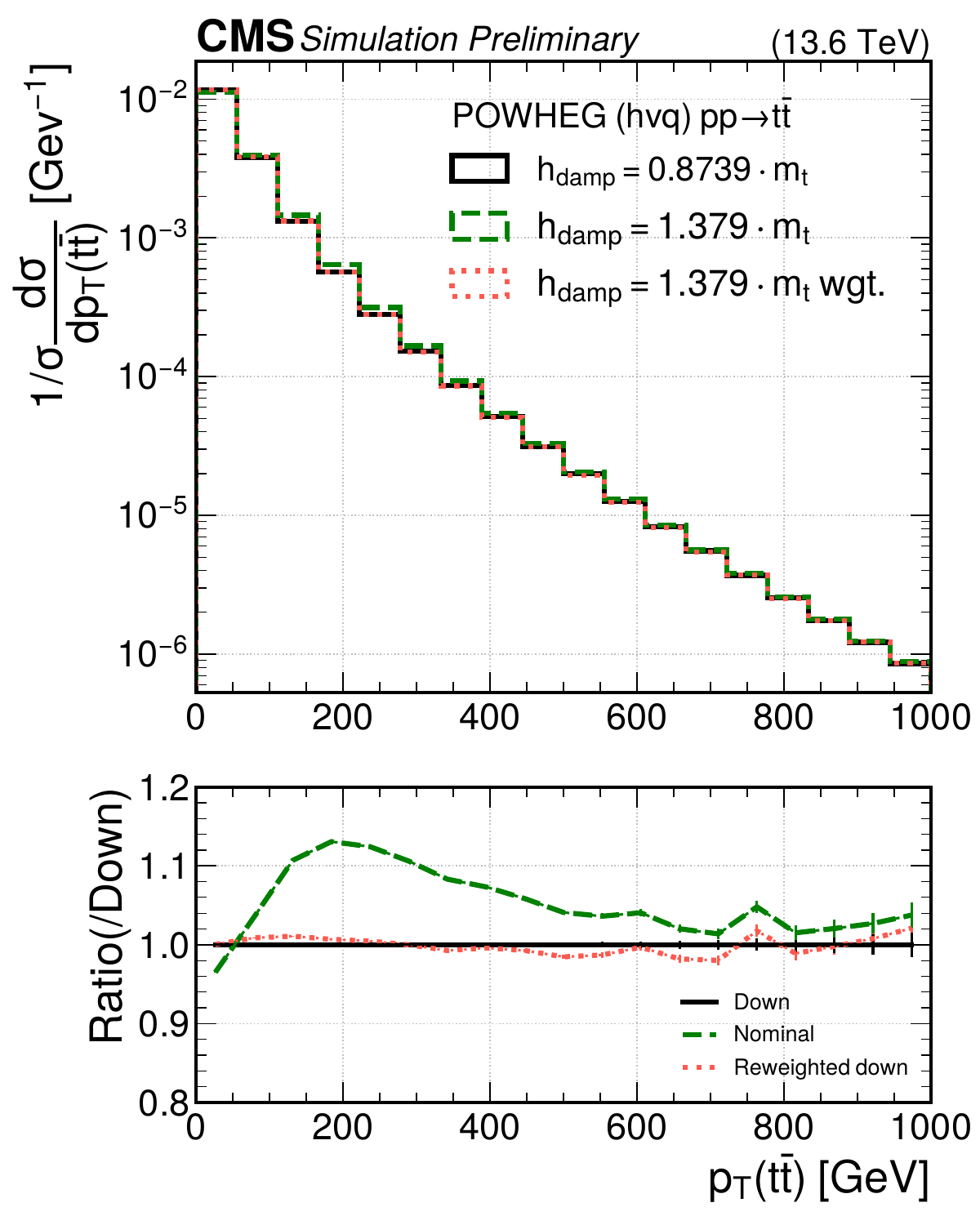}
\includegraphics[width=0.45\textwidth]{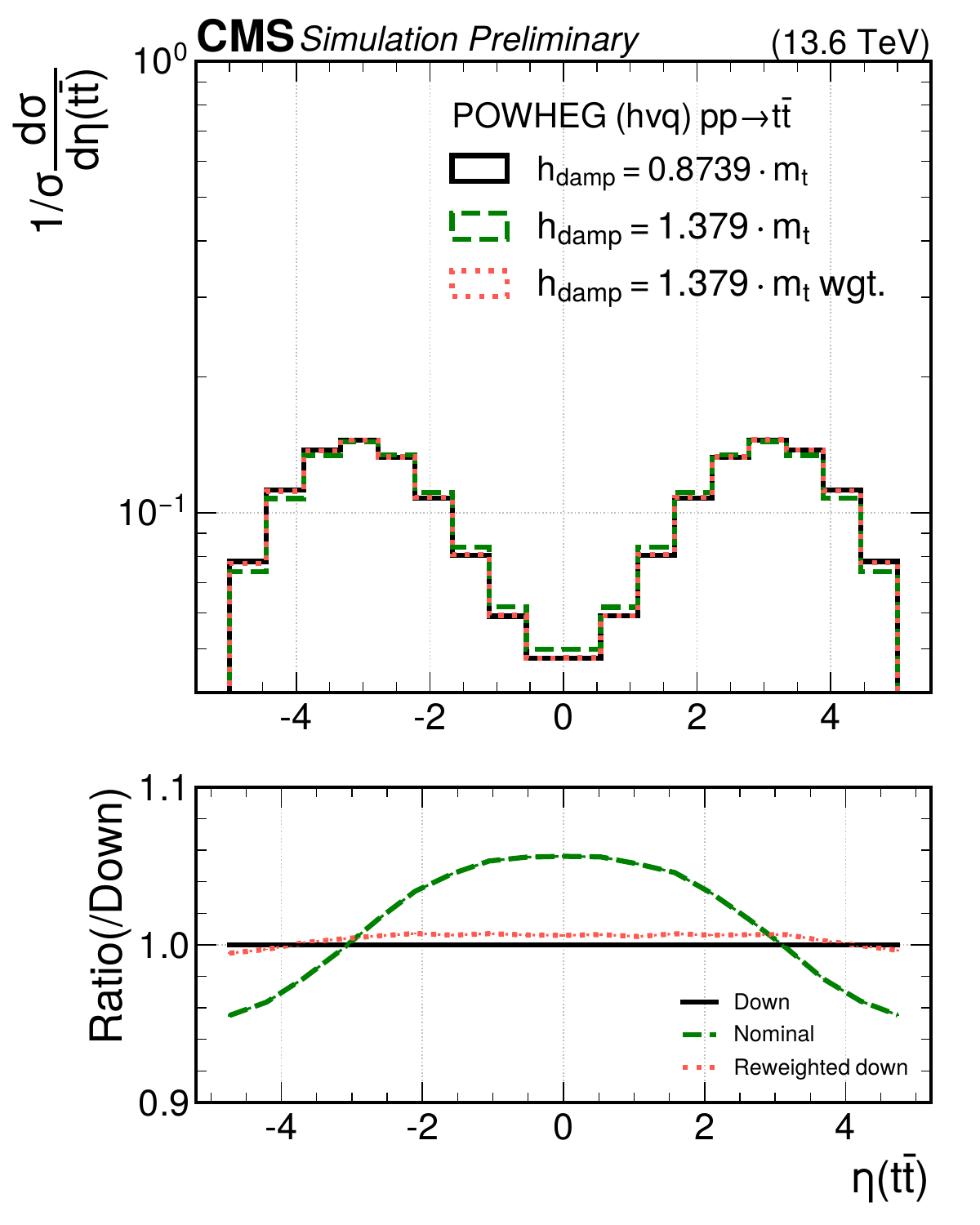}

\caption{The normalised differential cross section as a function of the $\pt$  (left) or $\eta$ (right) of the $\ttbar$  system at parton level for a MC simulated $\ttbar$  sample generated with the POWHEG hvq program, \cite{powheg1} \cite{powheg2}, for the CMS down variation of $\hdamp$ (0.8739 $\cdot$ $\mt)$ in black and the nominal value of $\hdamp$ (1.379 $\cdot$ $\mt$) in green. The red line shows the nominal sample reweighted to the down $\hdamp$ variation using the DCTR method \cite{DCTR}. The bottom pad shows the ratio of the nominal sample to the $\hdamp$ down variation before and after the reweighting. The error bars represent the statistical uncertainties due to the limited statistics of the MC.}
\label{fig:hdamp}
\end{figure}

The default MC sample of top pair production in CMS is generated using the Heavy Quark Process (HVQ)~\cite{hvq} of the event generator Powheg~\cite{powheg1}\cite{powheg2}. 
In Powheg, the resummation of the next-to-leading-order radiation is regulated by the $\hdamp$ variable, which enters the damping parameter D as in Eq.~\ref{eq:hdamp}:
\begin{equation}
    D = \dfrac{\hdamp^2}{\pt^{2} + \hdamp^2}
    \label{eq:hdamp}
\end{equation}
where $\pt$ is the transverse momentum of the particle and $\hdamp$ a parameter defined as $\hdamp = h\cdot \mt$, where $\mt$ is the mass of the quark top ($\mt$=172.5 GeV) and $h$ is a real number.
Since the parameter $\hdamp$ is not physical, an arbitrary value must be chosen in the simulation and varied to calculate the associated systematic uncertainty.
Two variations (down and up) from its nominal value are considered in CMS. 
The nominal value of $\hdamp$ is set to $1.379\cdot \mt$, while the down (up) variation is $0.8738 \cdot \mt$ ($2.305 \cdot \mt$).
Two separate neural network models are trained to reweight the nominal $\hdamp$ to these variations. 
For the training 40 million events are generated for each $\hdamp$ value, and parton-level information is used as input to the neural network. 
The Particle Flow Network (PFN) is employed for training, passing the quadrimomentum and the particle PID ($\pt$, $y$, $\phi$, $m$, PID) of the top and antitop to it.
The performance of reweighting is checked by applying the weights to the $\pt$ and $\eta$ observables of the $\ttbar$ system. 
Results, reported in Fig. \ref{fig:hdamp}, indicate that the reweighted and original samples agree with a method closure within 2$\%$.
In these proceedings, we present results for the down variation of $\hdamp$, but similar results are obtained for the up variation as reported in \cite{DPS}.

\begin{figure}[htbp]
\centering
\includegraphics[width=0.45\textwidth]{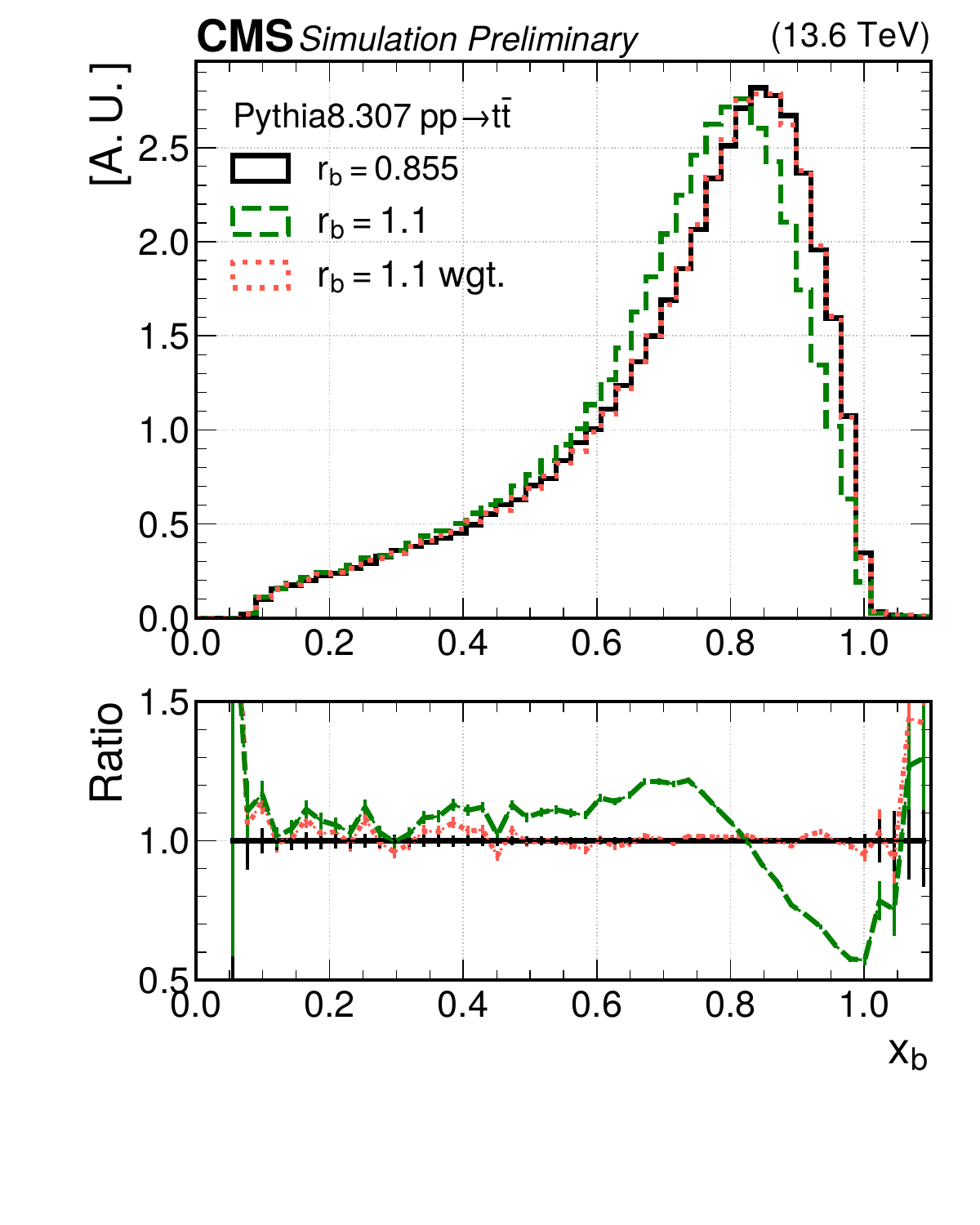}
\includegraphics[width=0.45\textwidth]{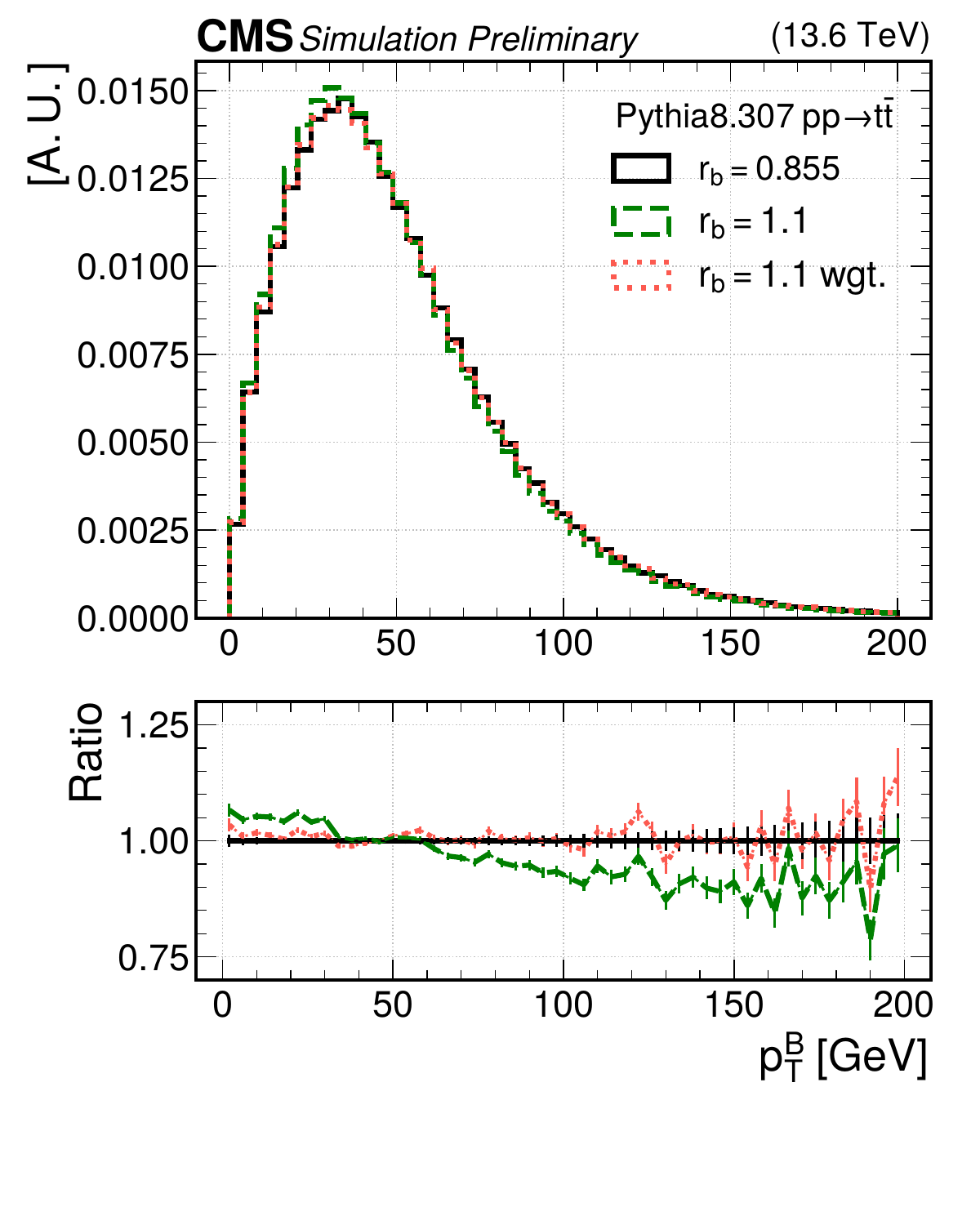}

\caption{The distribution of $\xb$ (left) or $\ptB$ of the B-hadron (right) of the $\ttbar$ system for a MC simulated $\ttbar$ sample generated with Pythia 8 \cite{pythia}, for the CMS nominal variation of $\rb$ (0.855) in black and a second value of $\rb$ (1.1) in green. The red line shows the  sample with the variation of $\rb$ reweighted to the sample generated with the nominal $\rb$ using the DCTR method, \cite{DCTR}. The bottom pad shows the ratio of the sample with the variation of $\rb$ to the sample generated with the nominal $\rb$ before and after the reweighting. The error bars represent the statistical uncertainties due to the limited statistics of the MC.}
\label{fig:bfragm}
\end{figure}

\section{B-fragmentation continous reweighting}
\label{sec:Bfragm}

Another significant source of uncertainty in top physics is the B-fragmentation.
It involves the decay of a top quark into a b quark and the subsequent formation of a B-hadron through a process described by the Lund string model in Pythia ~\cite{pythia}.
The probability distribution for heavy quarks during this process is given by Eq. \ref{eq:fB}: 

\begin{equation}
    f_B(z) \propto \dfrac{1}{z^{1+br_bm_b^2}}(1-z)^a exp(-bm_t^2/z)
    \label{eq:fB}
\end{equation}
where z is the quark longitudinal mometum, $\mt$ the top quark mass and a and b free parameters to be tuned to experimental data. 
For b quarks, the Bowler modification is also considered, involving the b quark mass $m_{b}$ and the $\rb$ parameter in Pythia.
To assess systematic uncertainties, CMS recommends varying the $\rb$ parameter during MC event generation. In this project, the DCTR technique is employed to estimate the uncertainty related to B-fragmentation, which affects physical observables like $\xb$ and $\ptB$ representing the energy fraction transferred from the b quark to the B-hadron and the transverse momentum of the B-hadron, respectively.
In this scenario, a continous reweighting with 10 different values of $\rb$ parameter in Pythia in the range [0.6, 1.4] has been performed training a single DNN model to reweight the samples generated with the different $\rb$ values to the nominal sample with $\rb$ (0.855).
For the nominal $\rb$ sample 5M events and 5k events for each $\rb$ variation sample are passed as inputs to the DNN for a total amount of 10M events to train.
The $\xb$ variable, defined in Eq \ref{eq:xb}, is passed as the only input to the DNN: 
\begin{equation}
    x_b = \dfrac{2p_B \cdot q}{m_t^2} \cdot \dfrac{1}{1-m_{w}^2/m_{t}^2}
\end{equation}
\label{eq:xb}
where $\pB$ and $q$ are the four vector of the B-hadron and of the top quark and $\mW$ and $\mt$ the W boson and top quark mass, respectively.
The performance of reweighting is checked by applying the weights obatined in the training to the
$\xb$ and $\ptB$ observables of the $\ttbar$ system.
The results for one of the 9 values of $\rb$ (1.1) trained are shown in Fig.~\ref{fig:hdamp}.
The original sample and the reweighted one agree also in this case with a method closure within 2$\%$.
Similar results are obtained for all the other 8 values of $\rb$ tested, as reported in \cite{DPS}. 

\section{Results and Conclusions}
In these proceedings, the DCTR method is evaluated in two different scenarios: discrete reweighting of the $\hdamp$ variation in Parton-level Powheg HVQ events and continuous reweighting of a b-fragmentation function parameter in particle-level Pythia events.
The method is found to work very well in both scenarios, with a method clusure within 2$\%$.
The method is implemented into CMS software framework for both cases.
This approach has the potential to be applied to various other intriguing cases in the field of top physics and, more broadly, in other physics areas.



\end{document}